    \documentstyle[eqsecnum,multicols,aps]{revtex}
    
\begin{document}
    \setlength{\topmargin}{-2ex}

    \draft

    \title{ The Consistency of  Causal Quantum Geometrodynamics and  Quantum
    Field Theory  }
    \author{N. Pinto-Neto\thanks{e-mail address: \tt nelsonpn@cbpf.br} and 
    E.
    Sergio
    Santini\thanks{e-mail address: \tt santini@cbpf.br}}
    \address{Centro Brasileiro de Pesquisas F\'{\i}sicas, \\
    Rua Dr.\ Xavier Sigaud 150, Urca 22290-180 -- Rio de Janeiro, RJ -- Brazil}
    \date{\today}
    \maketitle
    \begin{abstract}

    We consider quantum geometrodynamics and parametrized quantum field
    theories in the framework of the Bohm-de Broglie interpretation. In the
    first case, and following the lines of our previous work \cite{must},
    where a hamiltonian formalism for the bohmian trajectories was
    constructed, we show the consistency  of the theory for any quantum
    potential, completing the scenarios for canonical quantum cosmology
    presented there. In the latter case, we prove the consistency of scalar
    field theory in Minkowski spacetime for any quantum potential, and we
    show, using this alternative hamiltonian method, a concrete example
    where Lorentz invariance of individual events is broken.
    \end{abstract}
    \pacs{PACS numbers: 98.80.Hw, 04.60.Kz, 04.20.Cv, 3.70+k, 03.65.Bz}

    \begin{multicols}{2}

    \section{Introduction}

    The Copenhagen interpretation of quantum mechanics assumes the
    existence of a classical domain outside the observed quantum system.
    The necessity of a classical domain comes, according with Von Neumann's
    approach, from the way it solves the measurement problem (\cite{omn}):
    in order to explain why a unique eigenvalue of some observable is
    measured out of all possible outcomes, it is postulated that  the wave
    function suffers a real collapse, a process that cannot be described by
    the unitary Schr\"odinger evolution but must occur outside the quantum
    world, in a classical domain. In a quantum theory of the whole
    Universe, there is no place for a classical domain. Hence, the
    Copenhagen interpretation cannot be used in quantum cosmology.
    However, there are some alternative solutions to this quantum
    cosmological dilemma. One can say that the Schr\"{o}dinger evolution is
    an approximation of a more fundamental non-linear theory which can
    accomplish the collapse of the wave function \cite{rim,pen}, or that
    the collapse is effective but not real, in the sense that the branches
    related to unmeasured eigenvalues of the observable which appear in the
    splitting of a wave function in an impulsive measurement disappear from the
    observer but do not disappear from existence. In this second category
    we can cite the Many-Worlds Interpretation \cite{eve} and the Bohm-de
    Broglie (BdB) Interpretation \cite{bohm1}\cite{bohm2}\cite{hol}. In the
    former, all the possibilities in the splitting are actually realized.
    In each branch there is an observer with the knowledge of the
    corresponding eigenvalue of this branch, but she or he is not aware of
    the other observers and the other possibilities because the branches do
    not interfere. In the latter, a point-particle in configuration space
    describing the observed system and apparatus is supposed to exist,
    independently on any observations. It follows a trajectory in configuration
    space which is not the classical one due to the action of a quantum 
    potential
    which arises naturally from the Schr\"{o}dinger equation.
    In the splitting of the wave function in a measurement, this point
    particle will enter into one of the branches (which one depends on the
    initial position of the point particle before the measurement, which is
    unknown), and the other branches will be empty. It can be shown
    \cite{hol} that the empty waves can neither interact with other
    particles, nor with the point particle containing the apparatus.
    Hence, no observer can be aware of the other branches which are empty.
    Again we have an effective but not real collapse (the empty waves
    continue to exist), but now with no multiplication of observers. Of
    course these interpretations can be used in quantum cosmology.
    Schr\"{o}dinger  evolution is always valid, and there is no need of a
    classical domain outside the observed system.

    The BdB interpretation has been sucessfully applied to quantum
    minisuperspace models \cite{vink,bola1,kow,hor,bola2,fab,fab2} and even
    to full superspace models \cite{must} \cite{tese}. In the first case it
    was discussed the classical limit, the singularity problem, the
    cosmological constant problem, and the  time issue.  It was shown in
    scalar and radiation models for the matter content of the early
    universe that quantum effects driven by the quantum potential can
    inhibit the formation of a singularity by producing a repulsive quantum
    force that counteract the gravitational field, yielding inflation. The
    quantum universe usually reach the  classical limit for large scale
    factors.  However, it is possible to have small classical universes and
    large quantum ones:  it depends on the  state functional and on initial
    conditions \cite{fab}. It was shown that  the quantum evolution of
    homogeneous hypersurfaces form the same four-geometry independently on
    the choice of lapse function \cite{bola1}.

    For the case of a general superspace model  (i.e. the full theory)  a
    BdB picture of quantum geometrodynamics  was constructed \cite{must}.
    As the BdB interpretation admits the notion of quantum trajectories (in
    this case, the evolution of spacelike geometries in superspace, the
    space of all spacelike geometries), called bohmian trajectories, we
    were able to construct a hamiltonian formalism with constraints which
    generates such orbits. \footnote{It is important to stress that such
    hamiltonian formalism is relevant only for the subquantum world of the BdB
    interpretation, where trajectories are suposed to have objective
    reality. It has no meaning in interpretations where this subquantum
    world is absent, as in the Copenhaguen interpretation.} This hamiltonian
    formalism differs from the classical one only by the presence of the
    quantum potential in one of its terms which deviates the bohmian
    trajectory from the classical one. In this framework, it was possible
    to study the structures generated by the quantum evolution of spacelike
    3-geometries using the theory of constrained hamiltonians developed by
    Dirac \cite{dirac} for the case of geometrodynamics \cite{tei1,hkt}. In
    this way it was shown that, irrespective of any regularization and
    factor ordering of the Wheeler-De Witt equation, the BdB interpretation
    of quantum cosmology yields basically two possible scenarios:  In the
    first one, the constraints which generates the quantum bohmian
    evolution of the geometry of the spacelike hypersurfaces obey an
    specific algebra, the Dirac-Teitelboim's algebra  \cite{tei1}, which
    assures a consistent evolution that form a non degenerate 4-geometry
    with two possibilities:  the usual  hyperbolic classical spacetime and
    an euclidean spacetime where the change of signature from Lorentzian to
    Euclidean is driven by the quantum potential.  In the second scenario,
    the constraints are still conserved in time but they form an algebra
    different from the Dirac-Teitelboim's algebra yielding  a consistent
    quantum evolution that form a degenerate 4-geometry.  For example, in
    the case of real solutions of the Wheeler-DeWitt equation, we have a
    structure  satisfying the Carrol group   connected with the strong
    gravity limit. Another example with a non-local quantum potential was
    also studied.

    In that same paper \cite{must} we left open the possibility of an
    inconsistent evolution:   a complicated non-local quantum potential
    would avoid the closure of the algebra, whatever it would be
    (Dirac-Teitelboim or another alternative algebra), implying that the
    constraints are not conserved in time. As noted by Dirac, it would
    imply an inconsistent hamiltonian evolution of the spacelike
    geometries. Soon we realized that such possibility could also happen in
    quantum field theory.  If this would be the case, then the subquantum
    reality of the BdB interpretation would be imposing selection rules on
    possible quantum states which are absent in other interpretations,
    either for quantum geometrodynamics as for quantum field theory. This
    could be used to find ways to distinguish experimentally between
    interpretations and/or impose new boundary conditions which could be
    used in quantum cosmology. In the case of quantum field theory, which
    is more accessible to experimental investigations, if one realize in
    nature quantum states which presents this patological bohmian
    behaviour, it would be a serious drawback of the BdB interpretation. On
    the contrary, if one finds impossible to realize such states in
    practice, this would be a strong point in favour of the BdB
    interpretation because it would give a reason for this impossibility
    which is not present in other interpretations.

    In the present work we  show that this is not the case:  the algebra of
    the constraints is closed for any quantum potential when restricted to
    the quantum bohmian trajectories and hence there is no inconsistency in
    the sense of Dirac, neither for quantum geometrodynamics, nor for
    quantum field theory. Hence, the subquantum world of the Bohm-de
    Broglie interpretation is not imposing any restriction at all to the
    possible quantum states of a field theory: the admissible quantum
    states are the same as in any other interpretation.

    We begin the present paper with the application of the formalism
    developped in \cite{must} to a parametrized quantum field theory in
    flat background. We obtain analogous results as in quantum
    geometrodynamics, namely, the break of Dirac-Teitelboim's algebra and
    the consistency of the theory, in the sense of Dirac, for any quantum
    potential. In this case, the break of Dirac-Teitelboim's algebra  means
    a loss of Lorentz invariance of individual events, a result already
    known in the literature \cite{hol}\cite{bohm87} but which is presented
    here in a different way.  Note that this break of Lorentz invariance
    appears only at the level of individual quantum trajectories, the
    bohmian trajectories or the subquantum world, a notion which is absent
    in other interpretations of quantum mechanics.  \footnote{The study of
    individual events in quantum field theory was initiated already in an
    early paper by Bohm \cite{bohm2} and developed in much more detail in
    \cite{bohm87} and \cite{hol}. The loss of Lorentz invariance of
    individual events does not contradict relativity because all
    statistical predictions of the BdB interpretation are the same as the
    predictions of the Copenhagen interpretation, which are confirmed
    experimentally. The quantum trajectories which break Lorentz invariance
    in the BdB interpretation will not be accessible to observation as long
    as the quantum theory in its current form is valid. However, it is
    possible that quantum theory will fail in some new domain, unexplored
    until now. For instance, following Bohm \cite{bohm2}\cite{bohm87}, in
    an extension of quantum theory to include stochastic processes there
    will exist some relaxation time along which the probability density is
    approaching, but is not yet equal, to $\mid\psi\mid^2$. An experiment
    in times shorter than this relaxation time might reveal this
    discrepancy,  yielding in general results which are not Lorentz
    invariant. In such situation, relativity would hold only as a
    statistical approximation valid for  a distribution close to
    equilibrium in the stochastic process of the subquantum world that
    underly quantum mechanics.}

    The present paper is organized as follows: in the next section we
    revisit the parametrized scalar field theory in  Minkowski spacetime
    and we synthesize the Teitelboim's result about the spacetime structure
    reflected in the constraint's algebra. In section III we apply the
    Bohm-de Broglie interpretation to a parametrized scalar quantum field
    theory and we show that the bohmian evolution of the fields,
    irrespective to any regularization and factor ordering of the
    functional Schr\"odinger equation, can be obtained  from a specific
    hamiltonian which is, of course, different from the classical one. We
    prove that this hamiltonian formalism is consistent for any quantum
    potential. We then use this approach  to rederive, in a concrete
    example, the well known loss of Lorentz invariance of individual events
    in BdB theory in terms of the break of the Dirac-Teitelboim's  algebra
    of the constraints. In section IV we consider quantum geometrodynamics
    in the BdB interpretation following the approach of our previous work,
    and we prove the consistency of the theory for any quantum potential,
    completing the possible scenarios for the BdB view of quantum cosmology
    presented in \cite{must}. Section V is for  discussion and
    conclussions.  The appendix shows an alternative way to  compute a
    relevant Poisson bracket (PB).

    \section{Parametrized Field Theories}
    An essential  feature of geometrodynamics is the existence of  the
    super-hamiltonian
    and super-momentum constraints which are present due to the invariance
    of General Relativity (GR) under general coordinate transformations 
    \cite{rack}.
    We find  an analogous situations
    in systems with a finite number of  degrees of freedom and in field theory 
    in
    flat spacetime
    when  they are expresed as  {\it parametrized}  theories (see \cite{rack} 
    and
    \cite{lanckzos} cap VI). To parametrize a
    system with a finite number of  degrees of freedom, with  canonical 
    coordinates
    $X^i,(i=1...n)$ depending on the
    physical time $T$ and with lagrangean ${\cal L}_o$, we simply promote $T 
    \equiv
    X^0$ as being one of the canonical
    coordinates $X^\alpha=(T,X^i)$ (with $\alpha=0...n$) and
    depending on an arbitrary label
    time $t$: $T=T(t)$. In defining the conjugate momenta $\pi_T$ of $T$ a
    constraint appears: ${\cal H}\equiv
    \pi_T+H_o=0$ ( where $H_o\equiv\pi_i \frac{dX^i}{dT}- {\cal L}_{o}$ is the
    hamiltonian in the old
    coordinates
    and $\pi_i$ is the canonical  momentum conjugate to $X^i$ ).
    The original action functional
    $S=\int dT {\cal L}_{o} \biggr( T, X^i, \frac{dX^i}{dT} \biggl) = \int
    dT\biggr(\pi_i \frac{dX^i}{dT}- H_{o}\biggl)$
    will be expressed in the new canonical variables and after incorporating the
    constraint by mean a lagrange
    multiplier $N$, as
    \begin{equation}
    S=\int dt  \biggr( \pi_{\beta}\dot{X^{\beta}} - N{\cal H}  \biggl) ,
    \end{equation}
    where $X^0 = T$.
    This is the {\it parametrized} action and the variables $X^\alpha$ ,
    $\pi_\alpha$, $N $ are varied freely.

    For the case of a relativistic field theory we have  four parameters
    $X^\alpha$, instead of one $T$. We
    will introduce them as canonical coordinates, together with the fields
    $\phi(X^\alpha)$, by expressing them
    in terms
    of an arbitrary set of curvilinear coordinates $x^\beta$ as
    $X^\alpha=X^\alpha(x^\beta)$.
    In this manner (see below), we can build the field theory with the states
    defined on a general
    spacelike hypersurface, which plays the role of time. We  have a
    manifiest relativistic invariant hamiltonian formalism. The parametrized 
    form of
    the action
    of a field in
    flat spacetime will help us in the implementation of the BdB interpretation 
    to
    quantum gravity where
    the action is, from the beginning,  parametrized. In fact, up to now, it is
    imposible
    to deparametrize GR  in general by separating the dynamical (i.e relevant)
    degrees of freedom from the
    kinematical (i.e. redundant) ones. In GR we are forced to use redundant
    variables as canonical coordinates, and
    then constraints appears.

    Specifying, let $\phi(X^{\alpha})$  be a scalar field propagating in a
    $4$-dimensional flat spacetime  with
    minkowskian coordinates $ X^{\alpha} \equiv (T,X^{i})$. Greek indices run 
    from 0
    to 3 and latin indices from 1 to 3.
    Let us consider curvilinear coordinates  $ x^{\beta}=(t,x^{i})$ and a
    transformation
    \begin{equation}
    X^{\alpha}=X^{\alpha}(x^{\beta})
    \end{equation}
    For $t$ fixed, this equation defines a hypersurface with a spatial 
    coordinate
    system $x^{i}$ defined on it.
    We will have a family of hypersurfaces for different values of the parameter
    (label) $t$.
    The action in minkowskian coordinates is
    \begin{equation}
    S=\int d^{4}X {\cal L}_{o} \biggr(\phi,\frac{\partial \phi}{\partial
    X^{\alpha}}\biggl) ,
    \end{equation}
    where ${\cal L}_{o}$ stands for the lagrangean density in minkowskian
    coordinates. Writing the
    action in curvilinear coordinates we have:

    \end{multicols}
    \vspace{0.2cm}
    \ruleleft
    \vspace{0.2cm}
    \baselineskip=13pt

    \begin{equation}
    S=\int d^{4}x J  {\cal L}_{o} \biggr(\phi, \frac{\partial \phi}
    {\partial x^{\beta}}
    \frac{\partial x^{\beta}}{X^{\alpha}}\biggl)=\int d^{4}x {\cal 
    L}\biggr(\phi,
    \phi_{,i},
    \dot{\phi}, X^{\alpha}_{,i}, \dot{X^{\alpha}}\biggl) ,
    \end{equation}

    \vspace{0.2cm}
    \ruleright
    \begin{multicols}{2}
    \baselineskip=12pt

    \noindent
    where $\dot{\phi}\equiv\frac{\partial \phi}{\partial x^{0}}$,
    $\phi ,_{k} \equiv \frac{\partial}{\partial x^{k}} $, and

    \begin{equation}
    J \equiv \frac{\partial(X^{0}..X^{3})}{\partial(x^{0}..x^{3})}
    \end{equation}
    is the jacobian of the transformation. Here  ${\cal L}$ represents the
    lagrangean
    density in curvilinear coordinates. Defining the momentum $\pi_{\phi}$,
    conjugate to $\phi$, as usual,

    \begin{equation}
    \pi_{\phi}\equiv \frac{\partial {\cal L}}{\partial \dot{\phi}} ,
    \end{equation}
    we have the hamiltonian density

    \begin{equation}
    {\bf h} = \pi_{\phi} \dot{\phi}-{\cal L} ,
    \end{equation}
    yielding

    \begin{equation}
    {\bf h} = \frac{\partial x^{0}}{\partial X^{\alpha}} J T^{\alpha}_{\beta}
    \dot{X^{\beta}} \equiv K_{\beta}\dot{X^{\beta}} ,
    \end{equation}
    where $T^{\alpha}_{\beta}$ is the energy-momentum tensor of the field in
    minkowskian coordinates, given by

    \begin{equation}
    \label{tensorme}
    T^{\alpha}_{\beta} = \frac{\partial {\cal L}_{o}}{\partial
    \frac{\partial \phi}{\partial X^\alpha}} \frac{\partial \phi}{\partial
    X^\beta}-{\eta}^{\alpha}_\beta {\cal L}_{o} ,
    \end{equation}
    and $K_{\beta}$ is defined as

    \begin{equation}
    K_{\beta} \equiv \frac{\partial x^{0}}{\partial X^{\alpha}} J 
    T^{\alpha}_{\beta}
    \end{equation}

    The hamiltonian density ${\bf h}$ has a linear dependence in the 
    `kinematical
    velocities' $\dot{X^{\beta}}$
    because $K_{\beta}$ does not depend on them. The lagrangean density is given 
    by

    \begin{equation}
    {\cal L}=\pi_{\phi} \dot{\phi}-K_{\beta}\dot{X^{\beta}} ,
    \end{equation}
    We can define the `kinematical' momentum as
    \begin{equation}
    \Pi_{\alpha}\equiv\frac{\partial {\cal L}}{\partial \dot{X^{\alpha}}} =
    -K_{\alpha} ,
    \end{equation}
    which yields the constraint

    \begin{equation}
    \Pi_{\alpha}+K_{\alpha}=0 ,
    \end{equation}
    i.e.

    \begin{equation}
    \label{vice}
    {\cal H}_{\alpha} \equiv \Pi_{\alpha} + \frac{\partial x^{0}}{\partial
    X^{\beta}} J T^{\beta}_{\alpha} = 0 .
    \end{equation}
    In such a way, it is posible to write the action linearly in the dynamical
    velocities $\dot{\phi}$ and in the
    kinematical velocities $\dot{X^{\beta}}$

    \begin{equation}
    \label{alin}
    S=\int d^{4}x(\pi_{\phi} \dot{\phi}+\Pi_{\beta}\dot{X^{\beta}}) .
    \end{equation}
    In order to vary freely the action without worrying about the contraint
    (\ref{vice}), we must add the term
    $N^{\alpha}{\cal H}_{\alpha}$,  where $N^{\alpha}$ are lagrange multipliers,
    yielding

    \begin{equation}
    \label{Talin}
    S=\int d^{4}x(\pi_{\phi} 
    \dot{\phi}+\Pi_{\beta}\dot{X^{\beta}}-N^{\alpha}{\cal
    H}_{\alpha}) .
    \end{equation}

    It is possible to make a deeper analysis by projecting the constraints
    (\ref{vice})
    onto the normal direction
    and onto the paralell (or tangential) directions to the  hypersurfaces
    $t=constant$:

    \begin{equation}
    {\cal H}\equiv{\cal H}_{\alpha}n^{\alpha} ,
    \end{equation}
    \begin{equation}
    {\cal H}_{i}\equiv{\cal H}_{\alpha}X^{\alpha}_{i} ,
    \end{equation}
    where $X^{\alpha}_{i}$ are the components of the tangent vectors in the 
    basis
    $\frac{\partial}{\partial X^{\alpha}}$,
    $\frac{\partial}{\partial x^i}=\frac{\partial X^{\alpha}}{\partial
    x^i}\frac{\partial}{\partial X^{\alpha}}$ and
    the normal vector is defined by

    \begin{equation}
    \eta_{\alpha \beta}n^{\alpha}n^{\beta}=\epsilon=\mp 1 ,
    \end{equation}

    \begin{equation}
    n_{\alpha} X^{\alpha}_{i}=0 ,
    \end{equation}
    ($-$ for hyperbolic signature and $+$ for euclidean signature).
    Hence, the general form of the constraints is given by the sum of a 
    kinematical
    part plus a
    dynamical or field part:

    \begin{equation}\label{vinculop0}
    {\cal H} \equiv \Pi_{\alpha}n^{\alpha} +
    \frac{\partial x^{0}}{\partial X^{\beta}} J T^{\beta}_{\alpha}n^{\alpha} = 0 
    ,
    \end{equation}

    \begin{equation}\label{vinculopi}
    {\cal H}_{i} \equiv \Pi_{\alpha}X^{\alpha}_{i} + \frac{\partial 
    x^{0}}{\partial
    X^{\beta}
    } J T^{\beta}_{\alpha}X^{\alpha}_{i} = 0 .
    \end{equation}
    The constraint ${\cal H}$ is known as super-hamiltonian and the contraint 
    ${\cal
    H}_{i}$ as super-momentum.
    Expanding $N^{\alpha}$ in the basis $(n^{\alpha}, X^{\alpha}_{i})$, 
    $N^{\alpha}
    = Nn^{\alpha} + N^{i} X^{\alpha}_{i}$,
    we get the action in  parametrized form:

    \begin{equation}
    \label{ali}
    S=\int d^{4}x(\pi_{\phi} \dot{\phi} + \Pi_{\beta}\dot{X^{\beta}} - N{\cal H}
    - N^{i}{\cal H}_{i})
    \end{equation}

    The canonical variables $\phi, \pi_{\phi}, X^{\alpha}, \Pi_{\alpha}$ are 
    varied
    independently.
    The hamiltonian equations resulting from this variation will determine the
    evolution in time $t$
    of the canonical variables. Varying with respect to the lagrange multipliers 
    $N$
    and $N^i$, we obtain the
    constraints:

    \begin{equation}\label{VV}
    {\cal H}\approx 0, \hspace{0.5cm} {\cal H}_{i}\approx 0
    \end{equation}
    We write the last equations in Dirac's notation and terminology: the 
    constraints
    are {\it weakly}
    zero , which means that the Poisson brackets between  a
    functional of the canonical variables
    $A(\phi,\pi_{\phi},X^{\alpha}, \Pi_{\alpha})$  and a weakly zero constraint 
    are
    not
    necessarily zero
    \cite{dirac}. For consistency of
    the theory, the constraints  must be preserved in time, which
    means that their PB with the hamiltonian must be weakly zero. The 
    hamiltonian is

    \begin{equation}
    H = \int d^{3}x(N{\cal H} + N^{i}{\cal H}_{i})
    \end{equation}
    and the constraints ${\cal H}$ and
    ${\cal H}_i$ will be conserved in time only if all the PB between them,
    evaluated
    on two points
    $x$ e $y$ of the hypersurface, are weakly zero. Dirac made  this computation
    (with $\epsilon=-1$), and he showed that
    these brackets can
    be written as a linear combination of the original constraints (i.e. new
    constraints do not arise)
    satisfying the following algebra (known as `Dirac-Teitelboim's
    algebra')\footnote{It is not strictly an algebra because
    the structure constants depend
    on the metric \cite{hkt}}\cite{rack}\cite{dirac}:

    \begin{equation}
    \label{algebra1c}
    \{ {\cal H} (x), {\cal H} (y)\}={\cal H}^i(x) {\partial}_i \delta^3(x,y)-
    {\cal H}^i(y){\partial}_i \delta^3(y,x)
    \end{equation}
    \begin{equation}
    \label{algebra2c}
    \{{\cal H}_i(x),{\cal H}(y)\}={\cal H}(x) {\partial}_i \delta^3(x,y)
    \end{equation}
    \begin{equation}
    \label{algebra3c}
    \{{\cal H}_i(x),{\cal H}_j(y)\}={\cal H}_i(x) {\partial}_j \delta^3(x,y)-
    {\cal H}_j(y){\partial}_i \delta^3(y,x)
    \end{equation}
    where upper indices of the supermomentum are raised by the metric
    tensor $h_{ij}$ induced on the hypersurface
    $t=constant$, given by $h_{ij}=\eta_{\alpha
    \beta}X^{\alpha}_{,i}X^{\alpha}_{,j}$.
    (We take an usual convention in which we write the derivatives of the delta
    function with respect
    to the first argument every time: 
    $\delta_i(x,y)\equiv\frac{\partial}{\partial
    x^i}\delta(x,y)$
    and $\delta_i(y,x)\equiv\frac{\partial}{\partial y^i}\delta(y,x)$).
    Dirac found this result with the constraints given in the form  
    (\ref{vice}).

    \subsection{An important result by Teitelboim}
    It is appropiate at this point to remind a result that will be of 
    fundamental
    importance
    to our approach, which was obtained by Claudio Teitelboim \cite{tei1}. He
    obtained
    this algebra (but with the signature of spacetime appearing explicitily ) in 
    a
    general form
    that is independent of the form of the constraints and without assuming a
    Minkowski spacetime.
    He studied the deformations of  spacelike hypersurfaces embedded in a 
    riemannian
    spacetime. Intuitively, a labeled hypersurface can be deformed in general
    according to two operations:
    leaving it fixed in the embedding spacetime and relabeling its points or 
    keep
    fixed its labels and
    deform it into another  hypersurface.
    The first operation represents a deformation $\delta N^i \equiv \delta t 
    N^i$
    tangential to the
    hypersurface, generated by some $\bar{{\cal H}_{i}}$. The second operation
    represent
    a deformation $\delta N \equiv \delta t N$ ortogonal to the hypersurface,
    generated
    by some $\bar{{\cal H}}$. (For the previous case we have $\bar{{\cal
    H}_{i}}\equiv{\cal H}_{i}$ and
    $\bar{{\cal H}}\equiv {\cal H} $).
    Any functional $F$ of canonical variables (fields and kinematical variables)
    defined on the
    hypersurface will change, according to the hamiltonian given by
    \begin{equation}
    \label{hgm}
    \bar{H} = \int d^3x (N\bar{{\cal H}} + N^i\bar{{\cal H}}_i) \, ,
    \end{equation}
    in such a way that
    \begin{equation}
    \delta F = \int d^3x \{ F, \delta N \bar{{\cal H}} + \delta N^i\bar{{\cal 
    H}}_i
    \}\, ,
    \end{equation}
    which we write as
    \begin{equation}
    \label{hgmtei}
    \delta F = \int d^3x \{ F, \delta N^{\alpha} \bar{{\cal H}}_{\alpha} \}\, ,
    \end{equation}
    where $\bar{{\cal H}}_{0}\equiv\bar{{\cal H}}$ and $\delta N^{0}\equiv\delta 
    N
    $. Teitelboim follows
    a purely geometrical argument founded in the `path independence' of the
    dynamical evolution:
    the change in the canonical variables during the evolution from a given 
    initial
    hypersurface to a
    given final  hypersurface must be independent of the particular sequence of
    intermediary
    hypersurfaces along which the change is actually evaluated. Then, assuming 
    that
    the 3-geometries are
    embedded in a 4-dimensional non-degenerate manifold and consistency of the
    theory, he showed
    that the constraints $\bar{{\cal H}} \approx 0$ and $\bar{{\cal H}}_i
    \approx 0$ obey the following algebra (`Dirac-Teitelboim's algebra')


    \begin{equation}
    \label{algebra1m}
    \{ \bar{{\cal H}} (x), \bar{{\cal H}} (x')\}=-\epsilon[\bar{{\cal
    H}}^i(x) {\partial}_i \delta^3(x',x)
    -  \bar{{\cal H}}^i(x') {\partial}_i \delta^3(x',x)]
    \end{equation}
    \begin{equation}
    \label{algebra2m}
    \{\bar{{\cal H}}_i(x),\bar{{\cal H}}(x')\} = \bar{{\cal H}}(x)
    {\partial}_i \delta^3(x,x')  \, ,
    \end{equation}
    \begin{equation}
    \label{algebra3m}
    \{\bar{{\cal H}}_i(x),\bar{{\cal H}}_j(x')\} = \bar{{\cal H}}_i(x)
    {\partial}_j \delta^3(x,x')
    -\bar{{\cal H}}_j(x') {\partial}_i \delta^3(x',x) ,
    \end{equation}
    where indices of supermomentum are raised with the  metric $h_{ij}$ induced 
    on
    the
    hypersurface,
    $h_{ij}=g_{\alpha \beta}X^{\alpha}_{,i}X^{\alpha}_{,j}$, and $g_{\alpha 
    \beta}$
    is the metric
    of the embedding spacetime. The constant $\epsilon$ in  Eq.(\ref{algebra1m}) 
    can
    be  $\pm 1$
    depending if the 4-geometry where the hypersurfaces are embedded is 
    euclidean
    ($\epsilon = 1$) or
    hyperbolic ($\epsilon = -1$). This analysis can be applied to a field 
    evolving
    in a prescribed riemannian
    background or when the embedding spacetime is generated by the evolution, as 
    in
    GR. In the first
    situation (and in a flat background) the algebra imposes conditions for preserving the local Lorentz
    invariance. In the case of
    GR the algebra provides the conditions for the existence of spacetime: the
    evolution of a 3-geometry
    can be viewed as the `motion' of a 3-dimensional cut in a 4-dimensional
    spacetime (embeddability conditions). This result, when applied to the case
    of a parametrized field theory in a flat
    spacetime, means that the constraints obey the algebra given in
    (\ref{algebra1c})
    (\ref{algebra2c}) (\ref{algebra3c}).
    \subsection{A simple model}
    We will consider a scalar field in a flat spacetime, with a lagrangean given 
    by

    \begin{equation}
    {\cal L}_{o} =-\frac{1}{2}\biggr(\eta^{\alpha \beta}\frac{\partial
    \phi}{\partial X^\alpha}
    \frac{\partial \phi}{\partial X^\beta} + U(\phi)\biggl) \, ,
    \end{equation}
    where $\eta^{\alpha \beta}=\eta_{\alpha \beta}=diag(-1,1,1,1)$. Computing 
    the
    energy-momentum tensor given by
    Eq.(\ref{tensorme}), and substituting it in (\ref{vinculop0}) and
    (\ref{vinculopi}) we obtain the super-hamiltonian and
    super-momentum contraints as
    \begin{equation}
    \label{shc}
    {\cal H}=\frac{1}{\nu}(\Pi_{\alpha}\nu^{\alpha} + \frac{1}{2}\pi_{\phi}^{2} 
    +
    \frac{1}{2} \nu^2 (h^{ij}\phi_{,i}\phi_{,j} +  U(\phi))) = 0 \, ,
    \end{equation}

    \begin{equation}
    \label{supmc}
    {\cal H}_i= \Pi_ {\alpha} X^{\alpha}_{i} + \pi_\phi \phi_{,i} = 0   \, ,
    \end{equation}
    where the  vector normal to the hypersurface was written in the form (see
    \cite{lovelock} chap.7)
    $n^{\alpha}=\frac{\nu^{\alpha}}{\nu}$, where
    \begin{equation}
    \label{nu}
    \nu_{\alpha}\equiv -\frac{1}{3!}\epsilon_{\alpha \alpha1 \alpha2 \alpha 3}
    \frac{\partial ( X^{\alpha 1} X^{\alpha 2} X^{\alpha 3})}{\partial ( x^{1} 
    x^{2}
    x^{3})} \, ,
    \end{equation}
    and $\nu$ is the norm of $\nu^{\alpha}$

    \begin{equation}
    \nu= \sqrt{-\nu^\alpha \nu_{\alpha}} \, .
    \end{equation}
    It can be shown that $-\nu^\alpha \nu_{\alpha}=h$, where $h\equiv 
    det(h_{ij})$
    is the
    determinant of the  metric induced
    on the hypersurface.

    We know that the constraints obey the Dirac's algebra \cite{rack} (see the
    appendix for a
    prove  of (\ref{algebra1c}) in a somewhat different way). In the next 
    section
    we will quantize
    this model and we will interpret it within
    the Bohm-de Broglie picture.

    \section{ Parametrized Field Theory in the Bohm-de Broglie Interpretation}
    In this section we will study the Bohm-de Broglie interpretation of the
    parametrized field theory
    for the model presented in the last section.
    In first place, we quantize following Dirac prescription. Coordinates
    $\phi^A \equiv (X^{0}, X^{1}, X^{2}, X^{3}, \phi)$ and momentum
    $\pi_A \equiv (\Pi_{0}, \Pi_{1}, \Pi_{2}, \Pi_{3}, \pi_\phi)$ become
    operators, obeying commutation relations

    \begin{equation}
    [\phi^{A}(x),\phi^{B}(y)] = 0 , [\pi_{A}(x),\pi^{B}(y)] = 0 \, ,
    \end{equation}
    \begin{equation}
    [\phi^{A}(x),\pi_{B}(y)] = i \hbar \delta^A_B \delta(x,y) \, ,
    \end{equation}
    where $x, y$ are two points of the hypersurface. Constraints acts 
    anihilating
    the state,
    giving conditions for the possible states:
    \begin{equation}
    \label{smoc}
    \hat{{\cal H}}_i \mid \Psi  \! > = 0
    \end{equation}
    \begin{equation}
    \label{wdwc}
    \hat{{\cal H}} \mid \Psi  \! > = 0
    \end{equation}
    In the representation of `coordinates' $\phi^{A}(x)$, the  state of the 
    scalar
    field
    is given by the functional $\Psi[\phi^{A}(x)]$ and the momentum operator is
    a functional derivative: $\pi_{A}(x)=-i\hbar\frac{\delta}{\delta 
    \phi^{A}(x)}$.
    Substituting it into Eq. (\ref{smoc}), and taking into account
    the super-momentum Eq.(\ref{supmc}) we have:
    \begin{equation}
    \label{suminv}
    -i \hbar X^{\alpha}_{i} \frac{\delta \Psi}{\delta X^{\alpha}(x)} -
    i \hbar  \phi_{,i} \frac{\delta \Psi}{\delta \phi(x)} = 0 \, ,
    \end{equation}
    It  follows from this equation that   $\Psi$ is invariant under spatial
    coordinate transformations
    on the hypersurface.

    Using the super-hamiltonian given in Eq. (\ref{shc}), Eq. (\ref {wdwc})
    yields
    \end{multicols}
    \vspace{0.2cm}
    \ruleleft
    \vspace{0.2cm}
    \baselineskip=13pt

    \begin{equation}
    \label{supHpsi}
    {\cal H}(x)\Psi=\frac{1}{\nu}\biggr(-i\hbar \nu^{\alpha}\frac{\delta
    \Psi}{\delta X^{\alpha}(x) } -
    (\hbar)^2\frac{1}{2}\frac{\delta^2 \Psi}{\delta \phi(x)^2} +
    \frac{1}{2} \nu^2 \biggr( h^{ij}(x)\phi(x)_{,i}\phi(x)_{,j} +
    U(\phi(x))\biggl)\Psi \biggl) = 0 \, .
    \end{equation}

    \vspace{0.2cm}
    \ruleright
    \begin{multicols}{2}
    \baselineskip=12pt
    To interpret the above functional equation according to the BdB view, we 
    follow
    the usual procedure. First we write the wave functional
    in polar form $\Psi=A e^{\frac{i}{\hbar}S}$. Substituting it into Eq.
    (\ref{suminv}) yields two equations
    saying that  $S$ and $A$ are invariants under general space coordinate
    transformations:

    \begin{equation}
    \label{suminvS}
    X^{\alpha}_{i}\frac{\delta S}{\delta X^{\alpha}(x)} + \phi_{,i} \frac{\delta
    S}{\delta \phi(x)} = 0 \, .
    \end{equation}
    \begin{equation}
    \label{suminvA}
    X^{\alpha}_{i}\frac{\delta A}{\delta X^\alpha(x)} + \phi_{,i} \frac{\delta
    A}{\delta \phi(x)} = 0 \, .
    \end{equation}
    Substituting the polar form of $\Psi$ into Eq.(\ref{supHpsi}) we obtain two
    equations that
    depend on the factor ordering we choose. However, in any case,
    one of the equations will have the form, after dividing by $A$,

    \begin{equation}
    \label{hjc}
    \frac{1}{\nu}\biggr( \nu^{\alpha}\frac{\delta S}{\delta X^{\alpha}(x) } +
    \frac{1}{2}\biggr(\frac{\delta S}{\delta \phi }\biggl)^2
    +\frac{1}{2} \nu^2 W \biggl) +{\cal Q} = 0 \, .
    \end{equation}

    \noindent
    where, for simplicity of notation, we define
    $W \equiv h^{ij}(x)\phi(x)_{,i}\phi(x)_{,j} +  U(\phi(x))$.
    This  is like a Hamilton-Jacobi equation modified by the quantum potential,
    given by
    the last term. Only the quantum potential depends on  factor ordering and
    regularization;
    the other terms are well defined\footnote{To show explicitly this feature, we consider the hamiltonian 
with an arbitrary 
factor ordering scheme given by a differentiable functional $F$ with inverse
$F^{-1}$: 

  \begin{equation}
   {\cal H}=\frac{1}{\nu}(F\Pi_{\alpha}F^{-1}\nu^{\alpha} +
\frac{1}{2}F\pi_{\phi}F^{-1}\pi_{\phi}
   +
   \frac{1}{2} \nu^2 (h^{ij}\phi_{,i}\phi_{,j} +  U(\phi))) = 0 \, ,
   \end{equation}

 Writing this equation in the representation of `coordinates' $\phi^{A}(x)$,  we have

   \begin{equation}
   -i\hbar\frac{ \nu^{\alpha}}{\nu}\biggr( F\frac{\delta
   F^{-1}}{\delta X^{\alpha}(x)}\Psi+  \frac{\delta
   \Psi}{\delta X^{\alpha}(x) } \biggl) -
   (\hbar)^2\frac{1}{ 2 \nu}\biggr(F\frac{\delta
   F^{-1}}{\delta \phi(x) } \frac{\delta
   \Psi}{\delta \phi(x) }  + \frac{\delta^2 \Psi}{\delta \phi(x)^2} \biggl) +
   \frac{1}{2} \nu \biggr( h^{ij}(x)\phi(x)_{,i}\phi(x)_{,j} +
   U(\phi(x))\biggl)\Psi  = 0 \, .
   \end{equation}

    To interpret this functional equation according to BdB view we replace the
wave functional
 in the polar form
 $\Psi=A e^{\frac{i}{\hbar}S}$. The real part of the replaced equation yields,
after dividing by $A$ the following equation:

\begin{equation}
   \frac{ \nu^{\alpha}}{\nu}\frac{\delta S}{\delta X^{\alpha}(x) } +
   \frac{1}{2\nu}\biggr(\frac{\delta S}{\delta \phi }\biggl)^2
   +\frac{1}{2} \nu W  -\frac{\hbar^2}{2\nu}\frac{F}{A}\frac{\delta
F^{-1}}{\delta \phi}\frac{\delta A}{\delta \phi}-
\frac{\hbar^2}{2\nu A}\frac{\delta^2 A}{\delta \phi^2} = 0 \, .
   \end{equation}

   \noindent
   where, for simplicity of notation, we have  defined
   $W \equiv h^{ij}(x)\phi(x)_{,i}\phi(x)_{,j} +  U(\phi(x))$.

 As one can see, the sole modification one has is on the quantum potential. 
 The "classical" parts of this equation
 are not modified by  the introduction of $F$.}. All subsequent results depend only on the
    fact that ${\cal Q}$
    is a scalar density, which is true for any factor ordering, as it will be 
    shown
    shortly. According to the non-regulated version given in
    Eq (\ref{supHpsi}), ${\cal Q}$ is:

    \begin{equation}
    {\cal Q}=-\frac{1}{\nu}\frac{\hbar^2}{2A}\frac{\delta^2 A}{\delta \phi(x)^2} 
    \,
    .
    \end{equation}
    The other equation is:

    \begin{equation}
    \nu^{\alpha}\frac{\delta A^2}{\delta X^{\alpha}}+
    \frac{\delta(A^2 \frac{\delta S}{\delta \phi})}{\delta \phi}=0 \, .
    \end{equation}
    In the BdB interpretation the  canonical variables exist independently on
    measurements and  the evolution of canonical
    coordinates $\phi$ and $X^{\alpha}$ can be obtained from Bohm's guidance
    relations given by

    \begin{equation}
    \label{grx}
    \Pi_{\alpha} = \frac{\delta S(\phi, X^{\alpha})}{\delta X^{\alpha}} \, ,
    \end{equation}

    \begin{equation}
    \label{grc}
    \pi_\phi=\frac{\delta S(\phi, X^{\alpha})}{\delta \phi}  \, .
    \end{equation}
    Given the initial values of the field $\phi(t_0, x^i )$ and  kinematical
    variables
    $ X^{\alpha}(t_0, x^i)$ on a initial hypersurface $x^{0}=t=$const., we can
    integrate these
    first order differential equations and compute the bohmian trajectories, 
    i.e.,
    the values
    of the field $\phi(t,x^i)$ and  $X^{\alpha}(t,x^i)$ for any value of the
    parameter $t$.
    The evolution of those fields will be different from the classical one 
    because
    of the presence
    of the quantum potential in Eq. (\ref{hjc}).
    The classical limit is obtained by imposing the conditions for ${\cal Q}=0$.
    In this case, the functional $S$ obeys the classical Hamilton-Jacobi 
    equation
    and we know that integrating
    equations (\ref{grx}) and (\ref{grc}) the solutions obtained represent a
    classical field evolving in a
    Minkowski spacetime. This follows from the fact that the constraints of the
    classical theory
    satisfy the Dirac-Teitelboim's algebra (\ref{algebra1m}) (\ref{algebra2m})
    (\ref{algebra3m}) with $\epsilon=-1$.
    However, if the quantum potential is  different from zero, then $S$ is a
    solution of the {\it modified}
    Hamilton-Jacobi equation (\ref{hjc}). Hence, we cannot assert that the 
    obtained
    solution  for $\phi^A$
    still represent a field in a Minkowski spacetime. The quantum effects may 
    break
    Lorentz invariance
    and modify the einstenian causality of special relativity. Which type of
    structure
    corresponds to this case?  To answer this question we rewrite the  BdB  
    theory,
    originally
    formulated in a  Hamilton-Jacobi picture, in a
    hamiltonian picture.

    Using the Bohm's guidance relations (\ref{grx}) (\ref{grc}) we can write
    (\ref{hjc}) as:

    \begin{equation}
    \label{vshcn}
    \frac{1}{\nu}\biggr( \nu^{\alpha}\Pi_{\alpha} + \frac{1}{2} \pi_\phi{^2}
    +\frac{1}{2} \nu^2 W \biggl) + {\cal Q} = 0  \, .
    \end{equation}
    Note that, whatever is the form of the quantum potential  ${\cal Q}$,
    it must be a scalar density of weight one. This comes from the
    Hamilton-Jacobi equation Eq.(\ref{hjc}). From this equation we can express
    ${\cal Q}$ as

    \begin{equation}
    {\cal Q}= -\frac{1}{\nu}\biggr( \nu^{\alpha}\frac{\delta S}{\delta 
    X^{\alpha}(x)
    } + \frac{1}{2}(\frac{\delta S}{\delta \phi })^2
    +\frac{1}{2} \nu^2 W \biggl)
    \end{equation}

    \noindent
    We remember that $\nu=\sqrt{h}$ is a scalar density of weight 1 and that $S$ 
    is
    an invariant under
    general coordinate transformations on the hypersurface (this follows from 
    the
    supermomentum contraint
    applied to  $\Psi$, Eq.(\ref{suminvS})). Thus, $\frac{\delta S}{\delta
    X^{\alpha}}$ is a vector density
    which, when  contracted with the normal vector, produces a scalar density of
    weight 1. For the second
    term we use the same reasoning and the third term is obviously a scalar 
    density
    of weight 1. Hence
    ${\cal Q}$ is a sum of  scalar densities of weight 1.
    We can write Eq.(\ref{vshcn}) as

    \begin{equation}
    {\cal H} +  {\cal Q} = 0
    \end{equation}
    where ${\cal H}$ is the classical superhamiltonian given by (\ref{shc} ).The
    Bohm's quantum
    superhamiltonian  reads:

    \begin{equation}
    \label{hqc}
    {\cal H}_Q \equiv {\cal H} + {\cal Q}  \, .
    \end{equation}
    The hamiltonian that generates the bohmian trajectories, once the guidance
    relations (\ref{grx}) e (\ref{grc})
    are imposed initially, is

    \begin{equation}
    \label{hqca}
    H_Q = \int d^3x\biggr[N {\cal H}_Q + N^i{\cal H}_i\biggl] \, .
    \end{equation}
    This can be shown by noting that the guidance relations are consistent with 
    the
    constraints
    ${\cal H}_Q \approx 0$
    and ${\cal H}_i \approx 0$, because  $S$ satisfy (\ref{suminvS}) and
    (\ref{hjc}). Furthermore
    the guidance relations are conserved in the evolution given by the 
    hamiltonian
    (\ref{hqca}).
    This can be shown first by writing the guidance relations (\ref{grx})
    (\ref{grc})  in a form
    adapted for the hamiltonian formalism as:

    \begin{equation}
    \label{grxv}
    \Phi_\alpha \equiv \Pi_\alpha - \frac{\delta S}{\delta X^{\alpha}} \approx 0 
    \,
    ,
    \end{equation}

    \begin{equation}
    \label{grcv}
    \Phi_\phi \equiv \pi_\phi - \frac{\delta S}{\delta \phi} \approx 0 \, .
    \end{equation}
    Conservation in time of the guidance relations means that
    $\dot{\Phi}_{\phi} \equiv \{\Phi_{\phi}, H_Q \} = 0$ e $\dot{\Phi}_{\alpha}
    \equiv \{\Phi_{\alpha}, H_Q\}= 0$.
    This is equivalent to prove that their Poisson brackets with the constraints
    ${\cal H}_Q$ and ${\cal H}_i$
    are zero. Let us compute then $\{\Phi_\phi, {\cal H}_Q \}$, $\{\Phi_\alpha,
    {\cal H}_Q \}$, $\{\Phi_\phi, {\cal H}_i \}$
    and $\{\Phi_\alpha, {\cal H}_i \}$. The quantum superhamiltonian
    is given by

    \begin{equation}
    {\cal H}_Q \equiv {\cal H} + {\cal Q}=\frac{1}{\nu}\biggr(
    \nu^{\alpha}\Pi_{\alpha} +
    \frac{1}{2} \pi_\phi{^2}
    +\frac{1}{2} \nu^2 W \biggl) +{\cal Q}  \, .
    \end{equation}
    Computing, we have

    \end{multicols}
    \vspace{0.2cm}
    \ruleleft
    \vspace{0.2cm}
    \baselineskip=13pt

    \begin{eqnarray}
    \{\Phi_{\phi}(y), {\cal H}_Q(x) \}=\biggr\{ \Pi_\alpha - \frac{\delta 
    S}{\delta
    X^{\alpha}},
    \frac{1}{\nu}\biggr( \nu^{\alpha}\Pi_{\alpha} + \frac{1}{2} \pi_\phi{^2}
    +\frac{1}{2} \nu^2 W \biggl) +{\cal Q} \biggl\} = \nonumber \\
    -\frac{\delta}{\delta \phi(y)} \biggr\{\frac{1}{\nu}\biggr(
    \nu^{\alpha}\frac{\delta S}{\delta X^{\alpha}(x) } +
    \frac{1}{2}\biggr(\frac{\delta S}{\delta \phi }\biggl)^2
    +\frac{1}{2} \nu^2 W \biggl) +{\cal Q}\bigg\}-\frac{1}{\nu}\frac{\delta^2
    S}{\delta \phi^2}\Phi_{\phi} \, ,
    \end{eqnarray}

    \vspace{0.2cm}
    \ruleright
    \begin{multicols}{2}
    \baselineskip=12pt

    \noindent
    where the first term from the RHS of this equation stands for the functional
    derivative with respect to
    $\phi(y)$ of the LHS of the modified Hamilton-Jacobi equation, Eq 
    (\ref{hjc}).
    Hence, it is identically
    zero. The second term from RHS is weakly zero because of the guidance 
    relation
    (\ref{grcv}).
    Then, we have that

    \begin{equation}
    \{\Phi_\phi(y), {\cal H}_Q(x) \}=-\frac{1}{\nu}\frac{\delta^2 S}{\delta
    \phi(y)^2}\Phi_{\phi}(x)\approx 0 \, .
    \end{equation}
    For the bracket $\{\Phi_\alpha(y), {\cal H}_Q(x)\}$ we have

    \end{multicols}
    \vspace{0.2cm}
    \ruleleft
    \vspace{0.2cm}
    \baselineskip=13pt

    \begin{eqnarray}
    \{\Phi_\alpha(y), {\cal H}_Q(x)\}=-\frac{\delta }{\delta
    X^{\alpha}(y)}\biggr\{\frac{1}{\nu}\biggr( \nu^{\alpha}\frac{\delta 
    S}{\delta
    X^{\alpha}(x) }
    + \frac{1}{2}\biggr(\frac{\delta S}{\delta \phi(x)}\biggl)^2
    +\frac{1}{2} \nu^2 W \biggl) +{\cal Q} \biggl\} \nonumber \\
    -\frac{1}{\nu}\frac{\delta \nu^\beta}{\delta X^\alpha(y)}\Phi_\beta-
    \frac{\delta \nu^{-1}}{\delta X^{\alpha}(y)}\nu^{\beta}\Phi_{\beta}-
    \frac{1}{2}\frac{\delta \nu^{-1}}{\delta
    X^{\alpha}(y)}\biggr(\Phi_{\phi}^2+2\frac{\delta S}{\delta \phi}
    \Phi_{\phi}\biggl)
    -\frac{1}{\nu}\frac{\delta^2 S}{\delta\phi(x) \delta
    X^{\alpha}(y)}\Phi_{\phi}\approx 0 \, ,
    \end{eqnarray}

    \vspace{0.2cm}
    \ruleright
    \begin{multicols}{2}
    \baselineskip=12pt

    \noindent
    where the first term from the RHS of this equation stands for the functional
    derivative with respect to
    $\ X^{\alpha}(y)$ of the LHS of the modified Hamilton-Jacobi equation, Eq
    (\ref{hjc}). It is identically
    zero. The other terms are weakly zero because of the guidance relations
    (\ref{grxv}) (\ref{grcv}).
    To compute the Poisson brackets involving the supermomentum constraint, as
    $S$ is an invariant, then $\Phi_\alpha$ is a vector density and $\Phi_\phi$ 
    is a
    scalar density, both
    of weigth one. As ${\cal H}_i$ is the generator of space coordinate
    transformations, we get

    \begin{equation}
    \{\Phi_\phi(y), {\cal H}_i(x) \}=-\Phi_{\phi}(x) 
    \partial_{i}\delta(y,x)\approx
    0 \, ,
    \end{equation}

    \begin{equation}
    \{\Phi_\alpha(y), {\cal H}_i(x)
    \}=\Phi_i(x)\partial_{\alpha}\delta(y,x)-\Phi_\alpha(y)\partial_{i}\delta(y,x)\approx 0 \, .
    \end{equation}
    Combining these results we obtain

    \begin{equation}
    \dot{\Phi}_{\phi}=\{\Phi_\phi, H_Q\} \approx 0 \, ,
    \end{equation}

    \begin{equation}
    \dot{\Phi}_{\alpha,}=\{\Phi_\alpha, H_Q\} \approx 0 \, .
    \end{equation}
    Then, the Bohm's guidance relations are conserved.

    Knowing that the quantum potential does not depend on the momenta, we have 
    that
    the definitions
    of the momenta in terms of the velocities are the same as in the classical 
    case:
    \begin{equation}
    \dot{\phi}=\{\phi,H_Q\}=\{\phi,H\} \, ,
    \end{equation}

    \begin{equation}
    \dot{X^{\alpha}}=\{X^{\alpha},H_Q\}=\{X^{\alpha},H\} \, .
    \end{equation}

    We now have the BdB theory written in  hamiltonian form and we want to know
    which type of structure
    corresponds to the Bohmian evolution generated by the hamiltonian
    (\ref{hqca})\footnote{We would like to emphasize that the BdB interpretation 
    is
    only a different interpretation: its relevant hamiltonian operator as
    well as its  functional Schr\"{o}dinger's equation are the same as in
    any other interpretation. Hence, the eigenstates, including the ground
    state, and statistical predictions are the same, the difference
    residing in the interpretation, where trajectories independent of any
    observation (an uderlying objective reality) are suposed to exist.
    Hence, the {\it classical like} hamiltonian (not strictly classical
    because of the presence of the quantum potential, which has no parallel
    in classical physics) presented above is an extra structure meaningfull
    only within the BdB interpretation. Such {\it classical like}
    hamiltonian formalism has no place in the Copenhaguen interpretation,
    where only probabilities of measurement results have an objective
    reality: the subquantum world does not exist.}.
    The constraints ${\cal H}_i\approx0$ and ${\cal H}_Q\approx0$ must be 
    conserved
    in time for the consistency of
    the theory.
    This will be true only if all the PB between two of them, evaluated on two
    points $x$ and $y$ on the
    hypersurface, are weakly zero.
    In the context of the Teitelboim's work described in section II, let us  
    analyze
    the algebra satisfied
    by the constraints ${\cal H}_i\approx0$ and ${\cal H}_Q\approx0$. The PB
    $\{{\cal H}_i (x),{\cal H}_j (y)\}$ satisfies
    Eq. (\ref{algebra3c}) because ${\cal H}_i$  in  $H_Q$ defined by Eq.
    (\ref{hqca}) is the same as
    in the classical theory. In the same way, $\{{\cal H}_i (x),{\cal H}_Q 
    (y)\}$
    satisfies Eq. (\ref{algebra2c}) since
    ${\cal H}_i$ is the generator of space coordinate transformations, and 
    because
    ${\cal H}_Q$ is a scalar density of
    weigth 1.  What remains to be verified is
    if the PB $\{{\cal H}_Q (x),{\cal H}_Q (y)\}$ closes in the same way as in
    (\ref{algebra1c}).
    We will show that this bracket is weakly zero for any quantum potential 
    (note
    that it does not mean
    that it closes necessarily as in Dirac-Teitelboim's algebra Eq.
    \ref{algebra1c}).   This means that the theory
    is consistent for any ${\cal Q}$ and thus for any state. We have

    \begin{eqnarray}
    \label{cp1}
    \{{\cal H}_Q (x),{\cal H}_Q (y)\}=\{{\cal H}(x),{\cal H} (y)\} +
    \{{\cal H} (x),Q (y)\} + \nonumber \\
    \{ Q (x),{\cal H} (y)\} \, .
    \end{eqnarray}

    \noindent
    From equation (\ref{hjc}) we can write the quantum potential as:

    \begin{equation}
    {\cal Q}= -\frac{1}{\nu}\biggr( \nu^{\alpha}\frac{\delta S}{\delta 
    X^{\alpha}(x)
    } + \frac{1}{2}(\frac{\delta S}{\delta \phi })^2
    +\frac{1}{2} \nu^2 W \biggl) \, .
    \end{equation}
    Replacing the last equation into  Eq.(\ref{cp1}) and taking into account the
    Bohm's guidance relations given
    by (\ref{grxv}) and  (\ref{grcv}) we find that

    \end{multicols}
    \vspace{0.2cm}
    \ruleleft
    \vspace{0.2cm}
    \baselineskip=13pt

    \begin{eqnarray}
    &&\{{\cal H}_Q (x),{\cal H}_Q (y)\}=
    \frac{1}{\nu(x)\nu(y)}\biggr(\biggr(\frac{\delta S}{\delta \phi(y)}
    \frac{\delta^2 S}{\delta \phi(x) \delta \phi(y)}+ \nonumber \\
    &&\nu^{\alpha}(y)\frac{\delta^2 S}{\delta X^{\alpha}(y) \delta
    \phi(x)}\biggl)\Phi_{\phi}(x)-
    \biggr(\frac{\delta S}{\delta \phi(x)}
    \frac{\delta^2 S}{\delta \phi(y) \delta \phi(x)}+ \nonumber \\
    &&\nu^{\alpha}(x)\frac{\delta^2 S}{\delta X^{\alpha}(x) \delta
    \phi(y)}\biggl)\Phi_{\phi}(y) +
    \nu^{\alpha}(y)\frac{\delta \nu^{\beta}(x)}{\delta
    X^{\alpha}(y)}\Phi_{\beta}(y) -
    \nu^{\alpha}(x)\frac{\delta \nu^{\beta}(y)}
    {\delta X^{\alpha}(x)}\Phi_{\beta}(y)\biggl) \approx 0 \, ,
    \end{eqnarray}

    \vspace{0.2cm}
    \ruleright
    \begin{multicols}{2}
    \baselineskip=12pt
    The RHS of this equation is weakly zero in view of Bohm's guidance relations
    (\ref{grxv}) and (\ref{grcv}).

    Hence, we see that the Bohm-de Broglie interpretation of a field theory in
    Minkowski spacetime is a
    consistent theory. However, the constraint's algebra may not close  as
    Dirac-Teitelboim's algebra. It will
    depend on  the quantum potential. If $Q$ breaks the Dirac-Teitelboim's 
    algebra
    then, following Teitelboim's result,
    the structure of the background spacetime will be modified. This means that
    Lorentz invariance is broken.
    A similar situation is shown in quantum
    geometrodynamics where the quantum potential determine the quantum evolution 
    of
    the Universe \cite{must} \cite{tese}. We will now show that already the 
    ground
    state of the free scalar field
    produces a quantum potential that breaks the Dirac-Teitelboim's algebra and,
    consequently, Lorentz invariance.

    The wave functional for  the ground state of the free scalar field is given 
    by
    (\cite{hatfield} chap. 10):

    \begin{equation}
    \label{vacio}
    \Psi_0[\phi,T]=e^{-\frac{iE_{0}T}{\hbar}} \eta e^{-\int d^3X d^3Y
    \phi(X)g(X,Y)\phi(Y)}  \, ,
    \end{equation}
    where $E_0$ is the renormalized energy,

    \begin{equation}
    g(X,Y)=\frac{1}{2}\int \frac{d^3k}{(2\pi)^3} \omega_k e^{i k.(X-Y)} \, ,
    \end{equation}
    and $\omega_k=\hbar \sqrt{k^2+m^2}$. The quantity $\eta$ is a normalization
    factor which does not depend on $\phi$, and hence it is of no importance in
    the follwing calculations. We denote $ X\equiv \vec{X}$ and $k\equiv 
    \vec{k}$
    for the 3-vectors.
    Computing the amplitude from (\ref{vacio}) and using that

    \begin{equation}
    \label{pqc}
    Q(\phi) = -\hbar ^2 \frac{1}{2A} \int d^3 X \frac{\delta^2 A}
    {\delta \phi^2} \, ,
    \end{equation}
    the quantum potential reads

    \end{multicols}
    \vspace{0.2cm}
    \ruleleft
    \vspace{0.2cm}
    \baselineskip=13pt

    \begin{equation}
    Q=-\frac{1}{2}\int d^3X  \biggr( \int d^3Y \frac{d^3k}{(2\pi)^3} \omega_k
    \cos\{k.(X-Y)\} \phi(Y)\biggl)^2 +
    \frac{1}{2}\int d^3 X \int \frac{d^3k}{(2\pi)^3} \omega_k \, .
    \end{equation}

    \vspace{0.2cm}
    \ruleright
    \begin{multicols}{2}
    \baselineskip=12pt
    \noindent
    The last term is the zero-point energy which can be renormalized to
    $E_0$ by choosing, e.g., a normal factor ordering in Eq. (\ref{pqc}). 
    Anyway,
    it is not important for our calculations because it is independent of
    the field $\phi$ and hence its  functional derivative with respect to
    $\phi$ vanishes.

    We write the last equation, after an adequate renormalization of  the zero 
    point
    energy, as
    \begin{equation}
    Q=\int d^3X f( X^i, \phi) \, ,
    \end{equation}
    where $f$  depends on  $X^i$ as a function and depends on $\phi$ as a
    functional, and is given by
    \end{multicols}
    \vspace{0.2cm}
    \ruleleft
    \vspace{0.2cm}
    \baselineskip=13pt

    \begin{equation}
    f\equiv -\frac{1}{2}\biggr(\int d^3Y \frac{d^3k}{(2\pi)^3} \omega_k
    \cos\{k.(X-Y)\}
    \phi(Y)\biggl)^2  \, .
    \end{equation}

    \vspace{0.2cm}
    \ruleright
    \begin{multicols}{2}
    \baselineskip=12pt
    \noindent
    Writing it in hypersurface coordinates $x^i$ we have

    \begin{equation}
    Q=\int d^3x J f(X^i(x^j),\phi) \, ,
    \end{equation}
    where $J$ is the jacobian $ J=\frac{1}{3!}\epsilon_{ijk}\epsilon^{abc}
    \frac{\partial X^i}{\partial x^a}\frac{\partial X^j}{\partial 
    x^b}\frac{\partial
    X^k}{\partial x^c}$. The  quantum
    potential density
    ${\cal Q}$  that enters into the quantum superhamiltonian of Eq. (\ref{hqc})
    will be:

    \begin{equation}
    {\cal Q}= J f(X^i(x^j),\phi)
    \end{equation}
    Let us calculate the Poisson bracket $\{{\cal H}_Q (x),{\cal H}_Q (y)\}$. We
    have

    \end{multicols}
    \vspace{0.2cm}
    \ruleleft
    \vspace{0.2cm}
    \baselineskip=13pt

    \begin{eqnarray}
    \{{\cal H}_Q (x),{\cal H}_Q (y)\}=\{{\cal H}(x),{\cal H} (y)\} +
    \{{\cal H} (x),Q (y)\} + \{ Q (x),{\cal H} (y)\}= \nonumber \\
    {\cal H}^i(x) {\partial}_i \delta^3(x,y)-
    {\cal H}^i(y){\partial}_i \delta^3(y,x)+ \{{\cal H} (x),Q (y)\} + \{ Q 
    (x),{\cal
    H} (y)\} \, ,
    \end{eqnarray}

    \vspace{0.2cm}
    \ruleright
    \begin{multicols}{2}
    \baselineskip=12pt

    \noindent
    where the first two terms on the RHS are exactly those appearing in the 
    Dirac's
    algebra Eq.(\ref{algebra1c}).
    Hence, to maintain the  Dirac's algebra, it is necessary that
    $ \{{\cal H} (x),Q (y)\} + \{ Q (x),{\cal H} (y)\} = 0$ (strongly zero).
    Meanwhile,

    \end{multicols}
    \vspace{0.2cm}
    \ruleleft
    \vspace{0.2cm}
    \baselineskip=13pt

    \begin{eqnarray}
    \{{\cal H} (x),Q (y)\} + \{ Q (x),{\cal H} (y)\}=
    +2\frac{\nu^\alpha(y)}{\nu(y)}f(y)\epsilon_{\alpha j k}\epsilon^{abc}
    \frac{\partial X^j}{\partial y_b}\frac{\partial X^k}{\partial 
    y_c}\frac{\partial
    \delta(y,x)}{\partial y_a}+ \nonumber \\
    2\frac{J(y)}{\nu(x)}\pi_{\phi}B(y)\int\frac{d^3k}{(2\pi)^3} \omega_k
    \cos{k.(X(y)-x)} -
    2\frac{\nu^\alpha(x)}{\nu(x)}f(x)\epsilon_{\alpha j k}\epsilon^{abc}
    \frac{\partial X^j}{\partial x_b}\frac{\partial X^k}{\partial 
    x_c}\frac{\partial
    \delta(x,y)}{\partial x_a}+ \nonumber \\
    2\frac{J(x)}{\nu(y)}\pi_{\phi}(y)B(x)\int\frac{d^3k}{(2\pi)^3} \omega_k
    \cos{k.(X(x)-y)}  \, ,
    \end{eqnarray}

    \vspace{0.2cm}
    \ruleright
    \begin{multicols}{2}
    \baselineskip=12pt

    \noindent
    and the RHS of this equation is evidently strongly different from zero. 
    Hence
    the Dirac-Teitelboim's algebra
    is not satisfied in this particular example. According to  Ref.\cite{tei1}, 
    the
    bohmian trajectories
    are generating a new structure which does not correspond to a relativistic 
    field
    propagating in Minkowski spacetime. In other words, we have shown the 
    breaking
    of Lorentz invariance in terms
    of the breaking of the Dirac-Teitelboim's algebra of the constraints.
    This constitutes an alternative derivation of a known result of the BdB
    interpretation, which is explained
    in Refs. \cite{hol}\cite{bohm87}\footnote{It goes succintly in the following
    way: the equation for the field
    $\phi$ can be found by taking the functional derivative of the modified
    Hamilton-Jacobi
    equation of the field in the original Minkowski coordinates
    (see\cite{tese}\cite{hol}).  Using the Bohm guidance relations,
    it is possible to show that $\phi$
    satisfies
    \begin{equation}
    \label{eob2}
    -\nabla^2  \phi(X,T)+\frac{\partial^2}{\partial T^2}\phi(X,T)+ m^2\phi(X,T)
    =-\frac{\delta Q[\phi(X),T]}{\delta \phi(X)}|_{\phi(X) = \phi(X,T)}
    \end{equation}
    This is the quantum version of the classical wave equation:

    \begin{equation}
    \label{eoc}
    -\nabla^2  \phi(X,T)+\frac{\partial^2}{\partial T^2}\phi(X,T)+ m^2\phi(X,T) 
    = 0
    \, .
    \end{equation}
    The `quantum force' that appears  in the RHS of (\ref{eob2}) is  responsible 
    for
    all the quantum effects.
    For the ground state we have

    \begin{displaymath}
    -\nabla^2  \phi(X,T)+\frac{\partial^2}{\partial T^2}\phi(X,T) + m^2\phi(X,T)
    =(-\nabla^2 + m^2)\phi(X)|_{\phi(X) = \phi(X,T)} \, ,
    \end{displaymath}
    which is not a Lorentz invariant equation. We emphasize that $\phi(X,T)$  is 
    a
    c-number at each spacetime
    point. It is the eigenvalue of the Schr\"odinger field operator:
    $\hat{\phi(x)}\mid \Psi  \! >=\phi(x)\mid \Psi  \! >$, evaluated along a 
    system
    "trajectory"
    (the bohmian trajectory), a notion that has no meaning in the conventional
    interpretation. Equation (\ref{eob2}) belongs only to the subquantum world 
    of
    the BdB interpretation. It is
    important to the reader to not confuse
    $\phi(x,t)$ with the Heisenberg field operator $\hat{\phi}(x,t)$ which
    satisfies the usual Lorentz invariant
    Klein-Gordon equation \cite{hol}.}.

    This method will be useful for quantum geometrodynamics.
    We point out that relativistic invariance is  lost only
    at the level of individual events. However, the field properties, as we 
    know,
    are basically statistical and are contained
    in the expectation values of the operators

    \begin{equation}
      <  \Psi \mid \hat{A} \mid \Psi  > = \int 
    \Psi*[\phi](\hat{A}\Psi)[\phi]D\phi
    \end{equation}
    whose invariant character is maintained. As long as the invariance of the
    individual events can be broken
    in general, as we saw explicitly in the last example, special relativity is
    verified in the laboratory
    only statistically. Lorentz invariance is a statistical effect
    \cite{hol} \cite{bohm87}. We point out that the conclusions obtained in this
    work are easily extended
    to any Minkowski spacetime with dimension $n \geq 2$.

    \section{Quantum geometrodynamics}
    In  Ref. \cite{must} we applied the BdB interpretation to canonical quantum
    cosmology. Let's
    briefly summarize
    what was done there and the results obtained. The steps taken here are
    analogous to what was done in the previous section.

    We
    quantized General Relativity  with a
    minimally coupled scalar field in an arbitrary potential.
    All results remain essentially the same for any matter field which couples
    uniquely with the metric, not with their derivatives.
    The classical hamiltonian of GR with a scalar field is given by:
    \begin{equation}
    \label{hgr}
    H = \int d^3x(N{\cal H}+N^j{\cal H}_j)
    \end{equation}
    The lapse function $N$ and the shift function $N_j$ are the
    Lagrange multipliers of the { \it super-hamiltonian constraint}
    ${\cal H}\approx 0$ and the {\it super-momentum constraint}
    ${\cal H}^j \approx 0$,
    respectively. The constraints are present due to the invariance of GR under
    spacetime coordinate transformations.
    They are given by

    \begin{equation}
    \label{h0}
    {\cal H} = \kappa G_{ijkl}\Pi ^{ij}\Pi ^{kl} +
    \frac{1}{2}h^{-1/2}\Pi ^2 _{\phi} + V
    \end{equation}

    \begin{equation}
    \label{hi}
    {\cal H}_j = -2D_i\Pi ^i_j + \Pi _{\phi} \partial _j \phi .
    \end{equation}
    being   $V$  the classical potential given by
    \begin{equation}
    \label{v}
    V = h^{1/2}\biggr[-{\kappa}^{-1}(R^{(3)} - 2\Lambda)+
    \frac{1}{2}h^{ij}\partial _i \phi\partial _j \phi+
    U(\phi)\biggl] .
    \end{equation}
    In these equations, $h_{ij}$ is the metric of closed 3-dimensional
    spacelike hypersurfaces, and $\Pi ^{ij}$ is its canonical momentum
    given by
    \begin{equation}
    \label{ph}
    \Pi ^{ij} = - h^{1/2}(K^{ij}-h^{ij}K) =
    G^{ijkl}({\dot{h}}_{kl} -  D _k N_l - D _l N_k ),
    \end{equation}
    where
    \begin{equation}
    K_{ij} = -\frac{1}{2N} ({\dot{h}}_{ij} -  D _i N_j - D _j N_i ) ,
    \end{equation}
    is the extrinsic curvature of the hypersurfaces (indices are raisen and 
    lowered
    by the 3-metric $h_{ij}$ and its inverse $h^{ij}$). The canonical momentum
    of the scalar field is now
    \begin{equation}
    \label{pf}
    \Pi _{\phi} = \frac{h^{1/2}}{N}\biggr(\dot{\phi}-N^i \partial _i \phi 
    \biggl).
    \end{equation}
    The quantity $R^{(3)}$ is the
    intrinsic curvature of the hypersurfaces and $h$ is the determinant
    of $h_{ij}$.
    The quantities $G_{ijkl}$ and
    its inverse $G^{ijkl}$ ($G_{ijkl}G^{ijab}=\delta ^{ab}_{kl}$) are
    given by
    \begin{equation}
    \label{300}
    G^{ijkl}=\frac{1}{2}h^{1/2}(h^{ik}h^{jl}+h^{il}h^{jk}-2h^{ij}h^{kl}),
    \end{equation}
    \begin{equation}
    \label{301}
    G_{ijkl}=\frac{1}{2}h^{-1/2}(h_{ik}h_{jl}+h_{il}h_{jk}-h_{ij}h_{kl}),
    \end{equation}
    which is called the DeWitt metric. The quantity $D_i$ is the $i$-component
    of the covariant derivative operator on the hypersurface, and
    $\kappa = 16 \pi G/c^4$.
    The algebra of the constraints close in the following form
    \cite{hkt}\cite{dew}:

    \begin{eqnarray}\label{algebra}
    \{ {\cal H} (x), {\cal H} (x')\}&=&{\cal H}^i(x) {\partial}_i 
    \delta^3(x,x')-
    {\cal H}^i(x'){\partial}_i \delta^3(x',x) \nonumber \\
    \{{\cal H}_i(x),{\cal H}(x')\}&=&{\cal H}(x) {\partial}_i \delta^3(x,x')  \\
    \{{\cal H}_i(x),{\cal H}_j(x')\}&=&{\cal H}_i(x) {\partial}_j 
    \delta^3(x,x')+
    {\cal
    H}_j(x'){\partial}_i \delta^3(x,x') \nonumber
    \end{eqnarray}
    It is a feature of the hamiltonian of GR that the 4-metrics
    $ds^{2}=-(N^{2}-N^{i}N_{i})dt^{2}+2N_{i}dx^{i}dt+h_{ij}dx^{i}dx^{j}$
    constructed by integrating the Hamilton equations, with the same initial
    conditions, describe the
    same four-geometry for any choice of $N$ and $N^i$.

    After quantizing this theory following the Dirac procedure, we obtain the
    Wheeler-DeWitt equation and,
    in order to construct the BdB interpretation, the wave functional is written
    in polar form $\Psi = A \exp (iS/\hbar )$ yielding two equations
    for $A$ and $S$
    which  of course depend on the factor ordering we choose. However, in any 
    case,
    one of the equations will have the form

    \begin{equation}
    \label{hj}
    \kappa G_{ijkl}\frac{\delta S}{\delta h_{ij}}
    \frac{\delta S}{\delta h_{kl}}
    + \frac{1}{2}h^{-1/2} \biggr(\frac{\delta S}{\delta \phi}\biggl)^2
    +V+Q=0 ,
    \end{equation}
    where $V$ is the classical potential given in Eq. (\ref{v}).
    Contrary to the other terms in Eq. (\ref{hj}),
    which are already well defined because they do not need any regularization, 
    only
    the precise form of $Q$ depends on the regularization
    and factor ordering procedures which are prescribed for the Wheeler-DeWitt
    equation. Eq. (\ref{hj}) is
    like the Hamilton-Jacobi equation for GR,
    suplemented by an extra term $Q$, the quantum potential.
    The trajectories of the 3-metric and scalar field (which
    in this interpretation always exist by assumption)
    can be obtained from the guidance relations
    \begin{equation}
    \label{grh}
    \Pi ^{ij} = \frac{\delta S(h_{ab},\phi)}{\delta h_{ij}} ,
    \end{equation}
    \begin{equation}
    \label{grf}
    \Pi _{\phi} = \frac{\delta S(h_{ij},\phi)}{\delta \phi} ,
    \end{equation}
    These are first order
    differential equations which can be integrated to yield the 3-metric
    and scalar field for all values of the $t$ parameter. These solutions
    depend on the initial values of the 3-metric and scalar field at some
    initial hypersurface.  The evolution of these fields will of course be
    different from the classical one due to the presence of the quantum
    potential term $Q$ in Eq. (\ref{hj}). What kind of structure do we obtain 
    when
    we integrate this equations
    in the parameter $t$? Does this structure form a 4-dimensional
    geometry with a scalar field for any choice of the lapse and shift
    functions? Note that if the functional $S$ were a solution of the
    classical Hamilton-Jacobi equation, which does not contain the quantum
    potential term,
    then the answer would be in the affirmative because we would be in the
    scope of GR.

    In order to answer the questions formulated above, we  moved from
    this Hamilton-Jacobi picture of quantum geometrodynamics to a
    hamiltonian picture, where strong results concerning
    geometrodynamics  exist \cite{tei1}\cite{hkt}. In this way we have
    constructed a
    quantum geometrodynamical picture of
    the Bohm-de Broglie interpretation
    of canonical quantum gravity and we found that once the Bohm's guidance
    relations, given by Eqs.
    (\ref{grh}) and (\ref{grf}), are imposed initially, the bohmian trajectories
    will be generated
    by the hamiltonian, $H_Q$,
    given by

    \begin{equation}
    \label{hq}
    H_Q = \int d^3x\biggr[N({\cal H} + Q) + N^i{\cal H}_i\biggl]
    \end{equation}
    where we define
    \begin{equation}
    \label{hq0}
    {\cal H}_Q \equiv {\cal H} + Q .
    \end{equation}

    We have found that the bohmian evolution of the 3-geometries
    can yield, depending on the quantum potential ( i.e. on the wave 
    functional),
    basically
    two possibles  types of structures which are determined by the algebra
    satisfied by the constraints:

    {\bf A}. A consistent non-degenerate four geometry. In this scenario the
    quantum potential does not break the Dirac-Teitelboim's algebra, and the 
    most
    important quantum effect is the change of signature yielding
    an euclidean spacetime.

    {\bf B}. A consistent but degenerate four-geometry indicating the presence 
    of
    special
    vector fields and the breaking of the spacetime structure as a single
    entity. The constraints satisfy an algebra different from Dirac-Teitelboim's
    algebra.

    We left open the possibility of a third structure that would correspond to
    an inconsistent bohmian evolution in which the algebra of the constraints 
    does
    not close i.e.
    one of the PB would not be zero (the constraints would not be conserved in
    time). The relevant PB is $\{{\cal H}_Q (x),{\cal H}_Q (x')\}$ (the other 
    two PB
    are weakly zero for any
    quantum potential as was shown in Ref. \cite{must}). Some complicated 
    non-local
    quantum potential could
    make this PB weakly different from zero.
    In the present section we show
    that the PB $\{{\cal H}_Q (x),{\cal H}_Q (x')\}$, when restricted to the 
    bohmian
    trajectories,
    is weakly zero  for any quantum
    potential. Hence there is no inconsitency: quantum geometrodynamics in the
    Bohm-de Broglie
    interpretation is always
    a  consistent theory for any quantum state. In order to show this fact, we
    calculate

    \end{multicols}
    \vspace{0.2cm}
    \ruleleft
    \vspace{0.2cm}
    \baselineskip=13pt
    \begin{eqnarray}\label{pbqgd}
    \{ {\cal H}_Q (x), {\cal H}_Q (x')\} = \{{\cal H}(x), {\cal H}(x')\}
    -2 \kappa G_{abcd}(x)\Pi^{cd}(x)\frac{\delta {\cal Q}(x')}{\delta h_{ab}(x)}
    \nonumber \\
    +2 \kappa G_{abcd}(x')\Pi^{cd}(x')\frac{\delta {\cal Q}(x)}{\delta 
    h_{ab}(x')}-
    h^{-1/2}(x)\Pi_{\phi}(x)\frac{\delta {\cal Q}(x')}{\delta \phi(x)} +
    h^{-1/2}(x')\Pi_{\phi}(x')\frac{\delta {\cal Q}(x)}{\delta \phi(x')}\, .
    \end{eqnarray}
    \vspace{0.2cm}
    \ruleright
    \begin{multicols}{2}
    \baselineskip=12pt

    \noindent
    Our steps are similar to the ones followed in the previous section.
    We write the quantum potential as follows from the modified
    Hamilton-Jacobi equation (\ref{hj})

    \begin{equation}
    \label{hjerrata}
    {\cal Q} = -\kappa G_{ijkl}\frac{\delta S}{\delta h_{ij}}
    \frac{\delta S}{\delta h_{kl}}
    - \frac{1}{2}h^{-1/2} \biggr(\frac{\delta S}{\delta \phi}\biggl)^2 - V \, ,
    \end{equation}
    and replace it into (\ref{pbqgd}) yielding

    \end{multicols}
    \vspace{0.2cm}
    \ruleleft
    \vspace{0.2cm}
    \baselineskip=13pt

    \begin{eqnarray}
    \{ {\cal H}_Q (x), {\cal H}_Q (x')\} = \{{\cal H}(x), {\cal H}(x')\}
    +2 \kappa G_{abcd}(x)\Pi^{cd}(x)\frac{\delta V(x')}{\delta h_{ab}(x)} +
    h^{-1/2}(x)\Pi \frac{\delta V(x')}{\delta \phi(x)} \nonumber \\
    - \kappa G_{abcd}(x')\Pi^{cd}(x')\frac{\delta V(x)}{\delta
    h_{ab}(x')}-h^{-1/2}(x')\Pi \frac{\delta V(x)}{\delta \phi(x')}
    +4\kappa^2 G_{abcd}(x)\Pi^{cd}(x)\frac{\delta^2 S}{\delta h_{ab}(x)
    h_{ij}(x')}\frac{\delta S}{\delta h_{kl}(x')}G_{ijkl}(x') \nonumber \\
    +2\kappa G_{abcd}(x)\Pi^{cd}(x)h^{-1/2}(x')\frac{\delta S}{\delta
    \phi(x')}\frac{\delta^2 S}{\delta h_{ab}(x) \delta\phi(x')}
    +2\kappa h^{-1/2}(x)\Pi_{\phi}(x) G_{ijkl}(x')\frac{\delta^2 S}{\delta 
    \phi(x)
    \delta h_{ij}(x')}\frac{\delta S}{\delta h_{kl}(x')}  \nonumber \\
    +h^{-1/2}(x)h^{-1/2}(x')\Pi_{\phi}(x)\frac{\delta S}{\delta
    \phi(x')}\frac{\delta^2 S}{\delta \phi (x)\phi(x')}
    -(x \longleftrightarrow x')\, ,
    \end{eqnarray}

    \vspace{0.2cm}
    \ruleright
    \begin{multicols}{2}
    \baselineskip=12pt

    \noindent
    where $(x \longleftrightarrow x')$ means the same expression with $x$ and  
    $x'$
    interchanged. In the
    RHS,  the terms proportionals to $\delta^3(x',x)$ (these terms come from the
    functional derivatives
    $\frac{\delta G_{ijkl}(x')}{\delta h_{ab}(x)}$ and $ \frac{\delta
    h^{-1/2}(x')}{\delta h_{ab}(x)}$)
    were
    cancelled with the
    terms that come from the term $-(x \longleftrightarrow x')$. The four terms 
    that
    follows after
    $\{{\cal H}(x), {\cal H}(x')\}$ will produce exactly  $-\{{\cal H}(x), {\cal
    H}(x')\}$ and  they
    will be cancelled out.
    Substituting the momenta expressed according Bohm's guidance relations

    \begin{equation}
    \Pi ^{ij}(x)=  \Phi^{ij}(x) + \frac{\delta S}{\delta h_{ij}(x)} \, ,
    \end{equation}

    \begin{equation}
    \Pi_{\phi}(x)=\Phi_{\phi}(x)+\frac{\delta S}{\delta \phi(x)} \, ,
    \end{equation}
    it is easy to see, using the symmetry properties of $G_{ijkl}$, that all 
    terms
    that are not
    weakly zero will be cancelled by pairs. At last we have

    \end{multicols}
    \vspace{0.2cm}
    \ruleleft
    \vspace{0.2cm}
    \baselineskip=13pt

    \begin{eqnarray}
    \{ {\cal H}_Q (x), {\cal H}_Q (x')\} =
    +4\kappa^2 G_{abcd}(x)\Phi^{cd}(x)\frac{\delta^2 S}{\delta h_{ab}(x)
    h_{ij}(x')}\frac{\delta S}{\delta h_{kl}(x')}G_{ijkl}(x') \nonumber \\
    +2\kappa G_{abcd}(x)\Phi^{cd}(x)h^{-1/2}(x')\frac{\delta S}{\delta
    \phi(x')}\frac{\delta^2 S}{\delta h_{ab}(x) \delta\phi(x')}
    +2\kappa h^{-1/2}(x)\Phi_{\phi}(x) G_{ijkl}(x')\frac{\delta^2 S}{\delta 
    \phi(x)
    \delta h_{ij}(x')}\frac{\delta S}{\delta h_{kl}(x')}  \nonumber \\
    +h^{-1/2}(x)h^{-1/2}(x')\Phi_{\phi}(x)\frac{\delta S}{\delta
    \phi(x')}\frac{\delta^2 S}{\delta \phi (x)\phi(x')}
    -(x \longleftrightarrow x')\, .
    \end{eqnarray}
    \vspace{0.2cm}
    \ruleright
    \begin{multicols}{2}
    \baselineskip=12pt

    \noindent
    The RHS of this equation is weakly zero because of the Bohm's guidance 
    relations
    and, then

    \begin{equation}
    \{ {\cal H}_Q (x), {\cal H}_Q (x')\} \approx 0
    \end{equation}
    This prove the consistency. Note that it was very important  to use the 
    guidance
    relations to close the algebra.
    It means that the hamiltonian evolution with the quantum potential
    (\ref{hjerrata}) is consistent only when restricted to the bohmian 
    trajectories.
    For other trajectories, it may be inconsitent. This is an important remark
    on the BdB interpretation of canonical quantum cosmology, which
    sometimes is not noticed.
    
    We would like to remark that all these results were obtained
without assuming any particular factor ordering and regularization
of the Wheeler-DeWitt equation. In the canonical approach that 
we follow, the problems of renormalization and 
regularization appear in the Wheeler-DeWitt (WDW) equation because of the 
appearance of second order functional derivatives at the same point. 
Some ways to attack this problem are described in  \cite{reg} \cite{japa1}, 
and in all of them the kinetic part of the Hamilton-Jacobi equation 
associated with the WDW equation is not affected. As we said above, only 
the quantum potential term is affected. As our results 
are independent of the explicit form of the quantum potential, our 
results are not affected by renormalization and values of coupling.

    \section{Conclusions}

    We have studied a hamiltonian description of the BdB interpretation of
    quantum field theory and canonical quantum gravity.

    In the first case, we have examined certain state functionals where the
    quantum potential have such a form that the Dirac-Teitelboim's algebra
    of the classical constraints of the hamiltonian picture is broken, and
    the structure of the background spacetime becomes different from the
    Minkowski one.  This implies the break of Lorentz invariance of
    individual events. We exhibited a concrete example given by the ground
    state of a free scalar field presenting this property.  This is a well
    known result (see \cite{hol} \cite{bohm87}), but we have shown it in
    terms of the break of Dirac-Teitelboim's algebra. In this manner, the
    BdB view of quantum field theory expressed in a hamiltonian approach
    can give a nice picture of the loss of Lorentz invariance of the
    bohmian subquantum world:  when Dirac-Teitelboim's algebra is broken,
    then Lorentz invariance of individual events is lost. We should
    remember that the statistical properties of the quantum field, which
    are the same in the BdB interpretation as in the conventional one,
    continue to be Lorentz invariant.

    We have shown that both quantum field theory and canonical quantum
    gravity have a consistent bohmian hamiltonian formulation, in the
    sense of Dirac, for any quantum potential, i.e., any state functional,
    when restricted to the bohmian trajectories. In the case of canonical
    quantum gravity, this complete the quantum geometrodynamical picture of
    quantum cosmology in the BdB view which was constructed in our previous
    paper \cite{must}.

    The importance of this result can be seen as follows: supose there were
    states which had inconsistent quantum evolution in the BdB
    interpretation, in the sense we have described.  Then, the subquantum
    reality of the BdB interpretation would be imposing selection rules on
    possible quantum states which are absent in other interpretations. This
    could be used to find ways to distinguish experimentally between
    interpretations and/or impose new boundary conditions which could be
    used in quantum cosmology. However, we have shown that this subquantum
    world is not imposing any restriction to the possible quantum states of
    a field theory: the admissible quantum states are the same as in any
    other interpretation.

    \section*{ACKNOWLEDGEMENTS}

    We would like to thank
    {\it Conselho Nacional de Desenvolvimento Cient\'{\i}fico e Tecnol\'ogico}
    (CNPq) of Brazil
    and {\it Centro Latinoamericano de F\'{\i}sica} (CLAF) for financial 
    support.
    We would also like to thank `Pequeno Seminario' of CBPF's Cosmology Group 
    for
    useful discussions.
    \vspace{1.0cm}

    \appendix
    \section*{Computing $\{ {\cal H}(x),{\cal H}(y)\}$ for the parametrized 
    field
    theory}

    We present here the computation of $\{ {\cal H}(x),{\cal H}(y)\}$ for the
    parametrized
    field theory. Even though it is a well known result, we compute it in a new
    way. The superhamiltonian is given by (Eq. \ref{shc})
    \begin{eqnarray*}
    {\cal H}=\frac{1}{\nu}(\Pi_{\alpha}\nu^{\alpha} + \frac{1}{2}\pi_{\phi}^{2} 
    +
    \frac{1}{2} \nu^2 (h^{ij}\phi_{,i}\phi_{,j} +  U(\phi))) \, ,
    \end{eqnarray*}
    which we write, for simplicity, as

    \begin{equation}
    {\cal H}=\nu^{-1}\mbox{\sc h} \, ,
    \end{equation}
    where we define
    \begin{equation}
    \mbox{\sc h}\equiv\Pi_{\alpha}\nu^{\alpha} + \frac{1}{2}\pi_{\phi}^{2} +
    \frac{1}{2} \nu^2 (h^{ij}\phi_{,i}\phi_{,j} +  U(\phi)) \, .
    \end{equation}
    Then, we have

    \end{multicols}
    \vspace{0.2cm}
    \ruleleft
    \vspace{0.2cm}
    \baselineskip=13pt

    \begin{eqnarray}
    \{ {\cal H}(x),{\cal H}(y)\}=\frac{1}{\nu(x)}\frac{1}{\nu(y)}\{\mbox{\sc
    h}(x),\mbox{\sc h}(y)\}+
    \frac{1}{\nu(x)}\mbox{\sc h}(y)\{\mbox{\sc h}(x), \frac{1}{\nu(y)}\} +
    \frac{1}{\nu(y)}\mbox{\sc h}(x)\{\frac{1}{\nu(x)},\mbox{\sc h}(y)\} \, .
    \end{eqnarray}

    \vspace{0.2cm}
    \ruleright
    \begin{multicols}{2}
    \baselineskip=12pt

    \noindent
    Using the fact that
    $\frac{\delta \nu_{\alpha}(x)}{\delta X^{\beta}(y)}=-\frac{\delta
    \nu_{\beta}(x)}{\delta X^{\alpha}(y)}$,
    which comes from (\ref{nu}), and the basic properties  of the  
    $\delta(x,y)$, it
    is possible to show that
    each  of the  last two brackets of the RHS of this equation are identically
    zero. Next, using the same argument
    and the fact that the potential $U(\phi)$ does not contain derivatives of 
    the
    metric, the last equation becomes:

    \end{multicols}
    \vspace{0.2cm}
    \ruleleft
    \vspace{0.2cm}
    \baselineskip=13pt

    \begin{eqnarray}
    \{ {\cal H}(x),{\cal H}(y)\}=-\frac{1}{\nu^2(y)}\Pi_{\beta}(y)
    \nu^{\alpha}(y) \frac{\partial \nu^{\beta}(y)}{\partial X^{\alpha}_{i}(x)}
    \frac{\partial}{\partial y^i}\delta(y,x)
    +\frac{1}{\nu^2(x)}\Pi_{\alpha}(x)\nu^{\beta}(x)\frac{\partial
    \nu^{\alpha}(x)}{\partial X^{\beta}_{i}(y)}
    \frac{\partial}{\partial x^i}\delta(x,y) - \nonumber \\
    h^{ij}(y)\Pi_{\phi}(y)\frac{\partial \phi}{\partial y^j}
    \frac{\partial}{\partial y^i}\delta(y,x)+ 
    h^{ij}(x)\Pi_{\phi}(x)\frac{\partial
    \phi}{\partial x^j}
    \frac{\partial}{\partial x^i}\delta(x,y) \, .
    \end{eqnarray}

    \vspace{0.2cm}
    \ruleright
    \begin{multicols}{2}
    \baselineskip=12pt

    \noindent
    Finally, is possible to show that the first term of the RHS is equal to

    \begin{eqnarray}
    -h^{ij}(y)\Pi_{\alpha}(y)\frac{\partial X^{\alpha}}{\partial
    y^j}\frac{\partial}{\partial y^i}\delta(y,x) \, ,
    \end{eqnarray}
    and the second is equal to

    \begin{eqnarray}
    +h^{ij}(x)\Pi_{\alpha}(x)\frac{\partial X^{\alpha}}{\partial
    x^j}\frac{\partial}{\partial x^i}\delta(x,y) \, ,
    \end{eqnarray}
    where Eq. (\ref{nu}) was used again. Thus

    \end{multicols}
    \vspace{0.2cm}
    \ruleleft
    \vspace{0.2cm}
    \baselineskip=13pt

    \begin{eqnarray}
    \{ {\cal H}(x),{\cal H}(y)\}=h^{ij}(x)\biggr(\Pi_{\alpha}(x)\frac{\partial
    X^{\alpha}}{\partial x^j}+
    \Pi_{\phi}(x)\frac{\partial \phi}{\partial 
    x^j}\biggl)\frac{\partial}{\partial
    x^i}\delta(x,y)
    -h^{ij}(y)\biggr(\Pi_{\alpha}(y)\frac{\partial X^{\alpha}}{\partial y^j}+
    \Pi_{\phi}(y)\frac{\partial \phi}{\partial 
    y^j}\biggl)\frac{\partial}{\partial
    y^i}\delta(y,x) \, ,
    \end{eqnarray}

    \vspace{0.2cm}
    \ruleright
    \begin{multicols}{2}
    \baselineskip=12pt

    \noindent
    which means that

    \begin{eqnarray}
    \{ {\cal H}(x),{\cal H}(y)\}={\cal H}^i(x)\frac{\partial}{\partial
    x^i}\delta(x,y)-
    {\cal H}^i(y)\frac{\partial}{\partial y^i}\delta(y,x) \, .
    \end{eqnarray}
    which is Eq. (\ref{algebra1c}).


    \end{multicols}

    \end{document}